\documentclass[reprint,superscriptaddress,showpacs,preprintnumbers,amsmath,amssymb,aps,prl]{revtex4-1}
\usepackage[colorinlistoftodos]{todonotes} % package to include comments
\usepackage{graphicx}% Include figure files
\usepackage{color}% Include figure files
\usepackage{dcolumn}% Align table columns on decimal point
\usepackage{bm}% bold math
\usepackage{amssymb}   % for mathx
\usepackage{hyperref}% add hypertext capabilities
\usepackage{amsmath}

\begin{document}

\title{Plasma wakes driven by photon bursts via Compton scattering}

\author{F. Del Gaudio}
\email{fabrizio.gaudio@tecnico.ulisboa.pt}
\affiliation{GoLP/Instituto de Plasmas e Fus\~ao Nuclear, Instituto Superior T\'ecnico, Universidade de Lisboa, 1049-001 Lisbon, Portugal}

\author{R. A. Fonseca}
\affiliation{GoLP/Instituto de Plasmas e Fus\~ao Nuclear, Instituto Superior T\'ecnico, Universidade de Lisboa, 1049-001 Lisbon, Portugal}
\affiliation{DCTI/ISCTE Instituto Universit\'ario de Lisboa, 1649-026 Lisboa, Portugal}

\author{L. O. Silva }
\email{luis.silva@ist.utl.pt}
\affiliation{GoLP/Instituto de Plasmas e Fus\~ao Nuclear, I nstituto Superior T\'ecnico, Universidade de Lisboa, 1049-001 Lisbon, Portugal}

\author{T. Grismayer}
\email{thomas.grismayer@ist.utl.pt}
\affiliation{GoLP/Instituto de Plasmas e Fus\~ao Nuclear, Instituto Superior T\'ecnico, Universidade de Lisboa, 1049-001 Lisbon, Portugal}

\date{\today}

\begin{abstract}

Photon bursts with a wavelength smaller than the plasma inter-particle distance can drive plasma wakes via Compton scattering. We investigate this fundamental process analytically and numerically for different photon frequencies, photon flux, and plasma magnetization. Our results show that Langmuir and extraordinary modes are driven efficiently when the photon energy density lies above a certain threshold. The interaction of photon bursts with magnetized plasmas is of distinguished interest as the generated extraordinary modes can convert into pure electromagnetic waves at the plasma/vacuum boundary. This could possibly be a mechanism for the generation of radio waves in astrophysical scenarios in the presence of intense sources of high energy photons.

\end{abstract}

%\pacs{52.65.-y,52.35.Fp,78.70.-g,52.25.Os}

\maketitle
%\section*{ Introduction }

Electrons, positrons, ions, photons, and neutrinos can all drive wakes while propagating in plasma~\citep{Tajima_PRL_1979,Chen_PRL_1985,Esarey_IEEE_1996, Bingham_PLA_1996, Silva_PRL_1999, Jones_IEEE_1987}.
The streaming particles see the plasma as a dielectric medium~\citep{Krall_Trivelpiece_1973} and can excite plasma modes via the action of their effective ponderomotive force~\citep{Silva_PRE_1999}.
However, the description of plasmas as a dielectric media breaks down at scales where the notion of averaged fields loses its meaning. Intuitively, this scale should at least be the Debye length --- a more conservative estimate being the inter-particle distance. 
In the case of the electromagnetic fields, this corresponds to distinguishing the dressed photons from the non-dressed photons. The dressed photons acquire an effective mass due to the collective interaction with the plasma and propagate according to a dispersion relation \citep{Silva_PRE_1999}. On the contrary, for wavelengths scales smaller than the inter-particle distance, the collective behavior cannot emerge since the photon can only interact with one electron at a time. Dreicer~\citep{Dreicer_PoF_1964} and later Gould~\citep{Gould_AoP_1972} paved the way for a kinetic theory of plasmas which includes the full radiation field: the averaged field produced by the plasma in the fluid limit and the photon nature of the radiation. Discrete-particle effects, such as photon-electron scattering~\citep{Compton_PR_M1923}, explain the saturation properties of cyclotron radiation masers~\citep{Dreicer_PoF_1964}, the relaxation to a thermal equilibrium of a photon-electron gas~\citep{Kompaneets_JETP_1957,Peyraud_JP_1968}, or the Comptonization of the microwave background~\citep{Sunyaev_ARAA_1980}. The hindrance to these previous analytical studies lies in the treatment of electrons as free particles and the collective plasma dynamics are thus neglected. Frederiksen~\citep{Frederiksen_APJL_2008} was the first to lift the veil by pioneering particle-in-cell simulations coupled to a Monte Carlo Compton module. He observed with this novel numerical tool the formation of plasma wakefield structures driven by a broadband burst of gamma rays. In this Letter, we investigate theoretically and numerically the fundamental process of collective plasma wakes excitation by photon drivers. The interaction of the injected photons with the plasma is solely due to Compton photon-electron scattering. We would like to emphasize that this present work differs fundamentally from previous studies where exotic or non-conventional photon drivers, still interacting with the plasma via the ponderomotive force, such as X-ray pulses \citep{Wettervik_PoP_2018} or incoherent optical lasers \citep{Benedetti_PoP_2014} have been considered to drive wakefields. We explore different regimes according to the photon frequency, the photon flux, and the initial magnetization of the plasma. In the case of a magnetized plasma, the burst can excite plasma wakes (extraordinary mode) that can convert into radiation. We put in perspective the implications of this process in both laboratory experiments and high energy astrophysics.  We confirm our findings with plasma simulations performed with the particle-in-cell code OSIRIS~\citep{OSIRIS1,OSIRIS2}. OSIRIS has been enriched during the past years with several modules that allow exploring kinetic plasma physics in the regime where strong radiation \citep{Vranic_CPC_2016} and quantum electrodynamics processes become relevant \citep{Vranic_CPC_2015,Grismayer_PoP_2016,DelGaudio_PRAB_2019}. Recently, a Compton scattering module has been implemented to the OSIRIS framework \citep{DelGaudio_JPP_2020} in a similar fashion as the one of Frederiksen \citep{Frederiksen_APJL_2008}, and Haugboelle \citep{Haugboelle_PoP_2013}.

\begin{figure*}
\includegraphics[width=\linewidth]{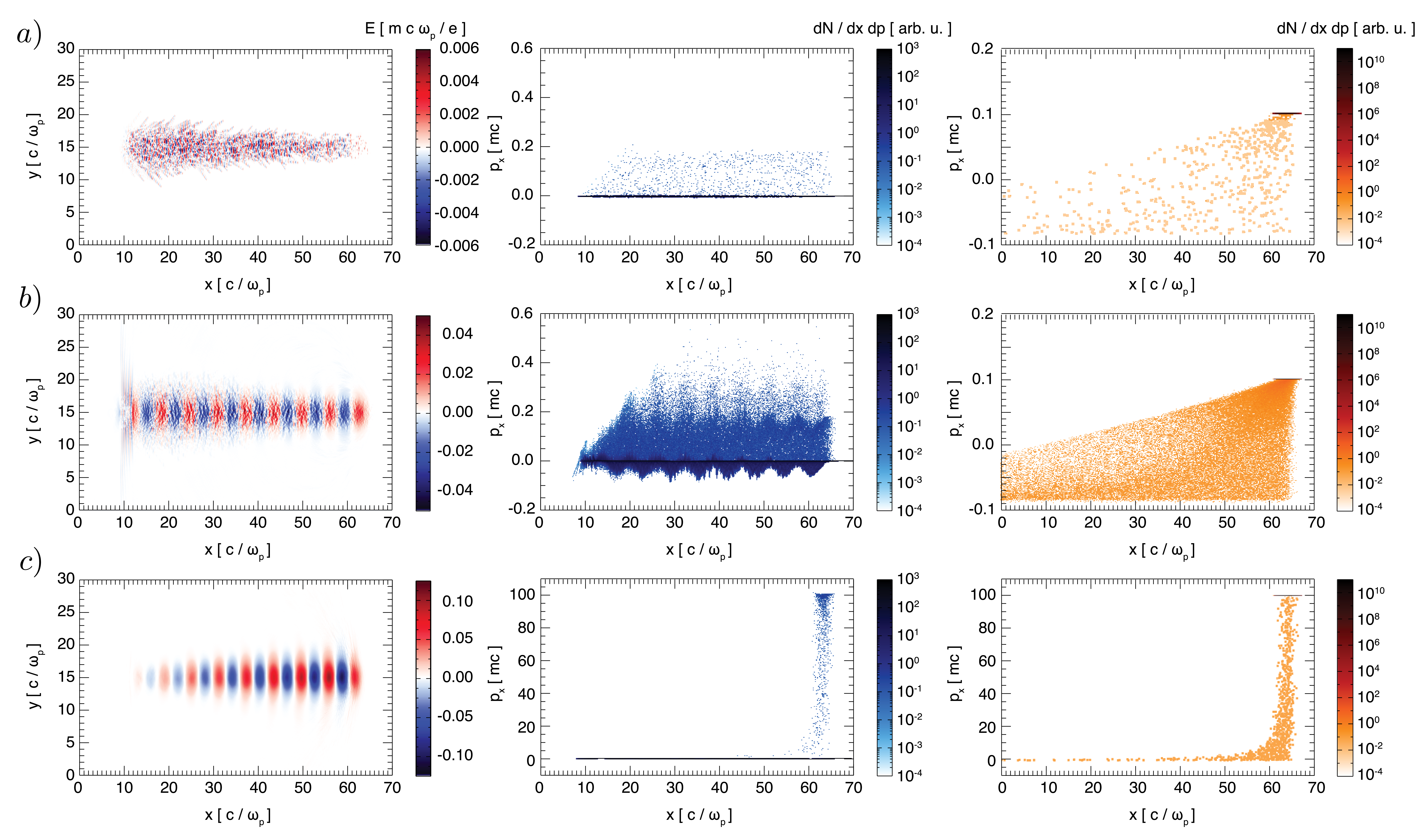} 
\caption{Different regimes of Compton-plasma interaction.
Left column: electrostatic field. Middle column: electron phase space. Right column: photon phase space.
a) incoherent wake: the space charge force is too weak to pull back most of the scattered electrons, $\lambda>\lambda_C$ and $\mathcal{E}_0 \sim 0.01~\mathcal{E}_{\mathrm{min}}$
b) coherent wake: the space charge is strong enough to pull back most of the electrons, $\lambda>\lambda_C$ and $\mathcal{E}_0 \sim \mathcal{E}_{\mathrm{min}}$
c) beam driven wake: the scattered electrons are relativistically kicked forward and formed on top of the photon burst a dense beam that contributes to drive the wake, $\lambda < \lambda_C$.}
\label{fig. wakes}
\end{figure*}

Let us consider a collimated photon burst of wavelength $\lambda = 2\pi c/\omega$ propagating in a cold unmagnetized plasma of density $n_p$ with $\hbar\omega$ the energy of the radiation quanta, and $c$ the speed of light. The Compton momentum exchange between an electron and a photon is, at the lowest order in $\lambda_C/\lambda$ ($\lambda_C = \hbar/mc$ is the Compton wavelength), $\Delta{\bf p} \simeq (\hbar\omega/c)({\bf n}_i-{\bf n}_f)$, where ${\bf n}_i$ and ${\bf n}_f$ are respectively the initial and the final direction of the photon. The angle averaged momentum exchange along the initial direction reads
\begin{equation}
\langle \Delta{p} \rangle  = \frac{1}{\sigma_T}\int_{-1}^{1}{\bf n}_i \cdot \Delta{\bf p} \frac{d\sigma}{d\mu}d\mu = \frac{\hbar\omega}{c}
\end{equation}
where $\mu = \cos{\theta}$, $\theta$ is the scattering angle, $\sigma_T$ the Thomson cross section, and $d\sigma/d\mu$ the differential scattering cross section in the Thomson limit. While the ions remain, in first approximation immobile, the electrons are scattered by the radiation burst of density $n_{\omega}$ at a rate $\nu_C = \sigma_Tn_{\omega}c$. The average force exerted on the electron fluid (at rest) by the photons is
\begin{equation}
\label{compton_force}
{\bf F}_C=\sigma_T n_{\omega}c\langle \Delta {\bf p}\rangle  = \sigma_T\mathcal{E}{\bf n}_i,
\end{equation}
where  $\mathcal{E} = n_{\omega}\hbar\omega$ is the radiation energy density. This result may appear as a special case obtained for a monochromatic burst. However, in the case of a broadband collimated burst, Peyraud \cite{Peyraud_JP_1968} has shown that the momentum exchange per unit of time (at the lowest order in $\lambda_C/\lambda$) on the electron fluid is $\sigma_T\int\hbar\omega N(\omega) d\omega$, which generalizes Eq.(\ref{compton_force}) since the integral term represents the energy density of the photons ($N(\omega)$ is the frequency distribution function of the photons). Hence a broadband burst containing the same energy density as a monochromatic burst results in the same average electron momentum $\langle \Delta p \rangle$.

We highlight that Eq.(\ref{compton_force}) corresponds to the average radiation reaction force \citep{LandauCTF} of an electron in the field of a EM plane wave of amplitude $E_w$ ($\mathcal{E} = E_w^2/4\pi$) \citep{classical_field_LL}. However, the quantum and classical description differ in nature, the first being discrete and stochastic, and the second continuous. In the special case of a (classical) electromagnetic pulse, the force on the electrons would show two components: the ponderomotive force ${\bf F}_P= -r_e \lambda^2 \nabla E_w^2$ ($r_e$ being the classical electron radius), and the radiation reaction force.We can compare the relative importance of the two components. For a pulse with a length on the order of the plasma wavelength $\lambda_p = 2\pi c/\omega_p$, $\vert \mathbf{F}_C\vert$ is much larger than $\vert \mathbf{F}_P\vert$ if 
\begin{equation}
\label{eq. length scale}
\frac{\lambda}{d_n} \ll  7.7  \left(\frac{r_e}{d_n}\right)^{1/4},
\end{equation}
where $d_n = n_p^{-1/3}$ is the inter-particle distance. For tenuous plasmas, the radiation reaction force exceeds the ponderomotive force when $\lambda \ll d_n$. Equation (\ref{eq. length scale}) can be rewritten in terms of photon energies as $\hbar\omega~\mathrm{[eV]}\gg 5.6 n_p^{1/4} ~\mathrm{[10^{17}~cm^{-3}]}$.

From a fundamental point of view, if a burst of photons with $\lambda \ll d_n$ enters a plasma, the concept of dielectric medium does not hold. Rather, a photon would see a very diluted ionized gas and would scatter with the electrons. The usual dispersion relation used for the propagation of light is not suitable in this situation and the photons travel at the speed of light between each collision.  

We show now that a burst of photons with $\lambda_C < \lambda < d_n$ that only interacts with the electrons through Compton scattering can excite some of the collective modes of the plasma. In the case of an unmagnetized plasma, this corresponds to the crossing of $\omega = kc$ with the Langmuir branch. We can use standard linear perturbation theory, with the introduction of the Compton force $\mathbf{F}_C$. The linearized equations of a one-dimensional cold fluid of plasma electrons are
\begin{eqnarray}
\partial_t n & = &-n_p\partial_{x}  v \nonumber \\
m\partial_t  v &=& -eE+\sigma_T\mathcal{E}  \\
\partial_{x} E &=& -4\pi e n, \nonumber
\end{eqnarray}
where $n$, $v$, and $E$ are the perturbed density, velocity, and field respectively. Solving for $E$, we obtain $\left(\partial_\xi^2 + 1\right) E = \sigma_T\mathcal{E}/e$, $\xi = \omega_p t -k_px$ is the frame of reference co-moving with the radiation burst, and $k_p=1/d_e$, $d_e = c/\omega_p$ the electron inertial length. 
We consider a driver of the form $\mathcal{E}(\xi) = \mathcal{E}_0\sin^2(\xi/2l)$, with $l = L/\lambda_p$ the normalized length of the burst. The wake amplitude at the back of the driver is obtained by convolution of the source term with the Green function of the harmonic operator
\begin{eqnarray}
\label{Eq:Efield_scaling}
\frac{e E(\xi)}{\sigma_T\mathcal{E}_0} &=& \frac{\sin(\pi l+\xi)\sin(\pi l)}{1-l^2}  \\ &=& 
\nonumber
\begin{cases}
l \ll  1 \rightarrow \pi l \sin(\xi) \\
l \simeq 1 \rightarrow (\pi/2)\sin(\xi) \\
l \gg 1 \rightarrow (\pi/2l^2) \sin(\xi).
\end{cases}\end{eqnarray}
Contrary to the wakes driven by lasers or electron beams \cite{Chen_PRL_1985,Esarey_IEEE_1996}, the amplitude of the electrostatic field does not depend on the plasma density but solely on the photon energy density \cite{field_amplitude}. However, we note that the optimal driver length corresponds to a resonant driver length, $L \sim \lambda_p$ as for  the standard laser case \cite{Tajima_PRL_1979}. For long symmetric drivers, the electrons experience multiple scattering, which tends to damp the amplitude of the electrostatic field. Although not demonstrated here, an asymmetric driver of arbitrary length, $L \gg \lambda_p$ with a rise/fall smaller than $\lambda_p$ still leads to a wake of reasonable amplitude $E \lesssim \sigma_T \mathcal{E}_0/e$. 

An upper threshold for the minimum photon energy density to observe a well-defined wake can be found by assuming that a significant fraction of the scattered electrons should be caught by the plasma wave oscillation. This amounts to say that $\langle \Delta{p} \rangle < eE/\omega_p$ or
\begin{equation}
\label{eq:coherence_scale}
\frac{\mathcal{E}_0}{n_pmc^2}>\frac{l_{\gamma}}{d_e}\frac{\hbar\omega}{mc^2}, 
\end{equation}
for a resonant driver, where $l_{\gamma} = 1/n_p\sigma_T$ is the photon mean free path. 
Equation({\ref{eq:coherence_scale}}) defines the minimum energy density $\mathcal{E}_{\mathrm{min}}$ to drive a well-defined wake. This condition is also equivalent to $\nu_C > \omega_p$ \cite{threshold}. An engineering formula for this minimum energy density in the case of a resonant driver is $\mathcal{E}_{\mathrm{min}}^{lab}  ~  \mathrm{[\mu J~\mu m^{-3}]}\sim n_p^{1/2}~\mathrm{[10^{14}cm^{-3}]~\hbar\omega~\mathrm{[eV]}}$ and in astrophysically relevant parameters
\begin{equation}
\label{eq:coherence_scale2}
\mathcal{E}_{\mathrm{min}}^{astro}  ~ \mathrm{[erg~cm^{-3}]}\sim 10^{6}~n_p^{1/2}~\mathrm{[cm^{-3}]}~\hbar\omega~\mathrm{[eV]}
\end{equation}
We would like to stress that the notion of a threshold as Eq.\ref{eq:coherence_scale} only applies when the driver cannot be described as a classical field \cite{classical_field_LL,LandauQED}. In other words, there is no such threshold for a laser driver since the momentum gain exerted by the radiation reaction force cannot be larger than the momentum associated to the wake. A plot summarizing the different regimes given by Eq.(\ref{eq. length scale}),  Eq.(\ref{eq:coherence_scale2}), and \cite{classical_field_LL} is presented in the Supplementary Material.

%\section*{ Simulations }

% ----- Figure two
\begin{figure}[t!]
\includegraphics[width=\linewidth]{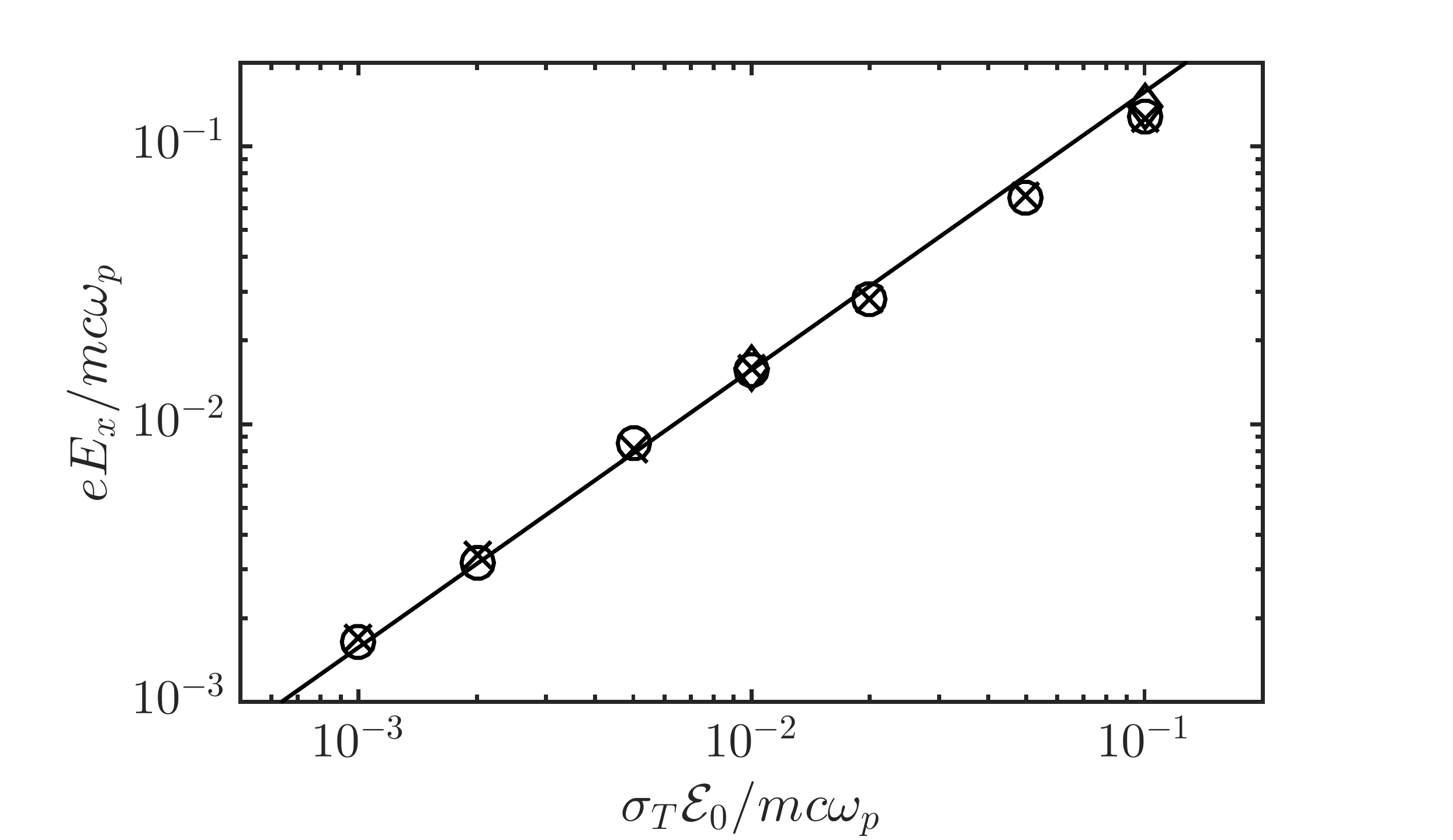}
\caption{Amplitude of the electrostatic field as a function of the energy density of a resonant photon burst. We simulated both a burst of mono-energetic photons, circles for $n_p=10^{18}~\mathrm{cm^{-3}}$ diamonds for $n_p=1~\mathrm{cm^{-3}}$, and crosses for a burst with an energy spread of $50\%$ ($n_p=10^{18}~\mathrm{cm^{-3}}$) . The linear theory of Eq.(\ref{Eq:Efield_scaling}) is displayed by the solid line.}
\label{fig. Amplitude}
\end{figure}

% ----- Simulations
We have carefully compared our analytical findings with 1D and 2D particle-in-cell simulations performed with OSIRIS-QED. The code includes a Compton scattering collision algorithm, which follows the pioneering work of Haugboelle ~\cite{Haugboelle_PoP_2013}. At each time-step, the number of macro-scattering is chosen by a Monte Carlo sampling of the Klein-Nishina cross section~\cite{Klein_ZP_1929}. The momentum of the scattered macro-particles is then updated following the wavelength shift of Compton scattering~\cite{Compton_PR_M1923}. We performed all of our simulations for a reference plasma density of $n_p = 1~\mathrm{cm^{-3}}$, and $n_p = 10^{18}~\mathrm{cm^{-3}}$.

A series of 2D simulations were conducted with a resonant radiation driver ($L_x = L_y = \lambda_p$). The 2D simulations have been performed in a domain of $72~d_e\times 30~d_e$ with $\Delta x=\Delta y=0.1~d_e$ and $\Delta t =0.07~\omega_p^{-1}$, with $16$ macro-particles per cell. Figure~\ref{fig. wakes}-a) $\lambda>\lambda_C$ ($\hbar\omega = 50$ keV) and $\mathcal{E}_0/\mathcal{E}_{\mathrm{min}}\sim 0.01$ corresponds to the regime where the photon energy density is below the threshold. Most of the scattered electrons cannot be pulled back by the space charge force and stream through the plasma. Physically, each streaming electron drives its own wake, creating electrostatic fluctuations \cite{Krall_Trivelpiece_1973}. The phase space shows that the electrons are also heated up. The maximum momentum of the electron population corresponds to the maximum momentum exchange with the photon, $p_{\mathrm{max}} = 2\hbar\omega/c$. When the photon energy density is around the threshold $\mathcal{E}_0/\mathcal{E}_{\mathrm{min}}\sim 1$, the linear wake is visible as displayed in Fig~\ref{fig. wakes}-b). The electron phase space exhibits the presence of plasma oscillations on top of the temperature induced by the Compton collisions. Note that the maximum longitudinal momentum is higher than in the previous case due to electron acceleration in the wake. The initial photon burst is very slightly depleted --- the scattered photons are slowly sliding back in the plasma as shown in the phase space. 

Figure~\ref{fig. wakes}-c) shows a regime, $\lambda<\lambda_C$ ($\hbar\omega = 50$ MeV), that goes beyond the exposed theory. When $\lambda<\lambda_C$, the angular cross section becomes relativistically beamed and each scattering propels the electron forward close to the speed of light as seen in the phase space. As a result, the photon burst becomes rapidly loaded with an increasing dense relativistic electron beam. In this regime, the beam is driving the wake \cite{Chen_PRL_1985}. The case of Frederiksen \cite{Frederiksen_APJL_2008} due to a long burst with a broad spectrum is hard to be characterized since both of the last aforementioned regimes are competing. We also verified the scaling of Eq.(\ref{Eq:Efield_scaling}) for a fixed plasma density with 1D simulations shown in Fig.~\ref{fig. Amplitude}. The 1D simulations have been performed with a moving window (which follows the driver pulse) $24~d_e$ long with $\Delta x_1=0.01~d_e$ and $\Delta t =0.0099~\omega_p^{-1}$, with $256$ macro-particles per cell. The length of the radiation burst is $\lambda_p$ and the energy of each photon is in the range $\hbar\omega=5-50$ keV, such that Eq.(\ref{eq:coherence_scale}) is fulfilled. We have also considered a beam with a $50\%$ spread in energy. As predicted, the spread does not influence the wake amplitude.

\begin{figure}[t]
\includegraphics[width=\linewidth]{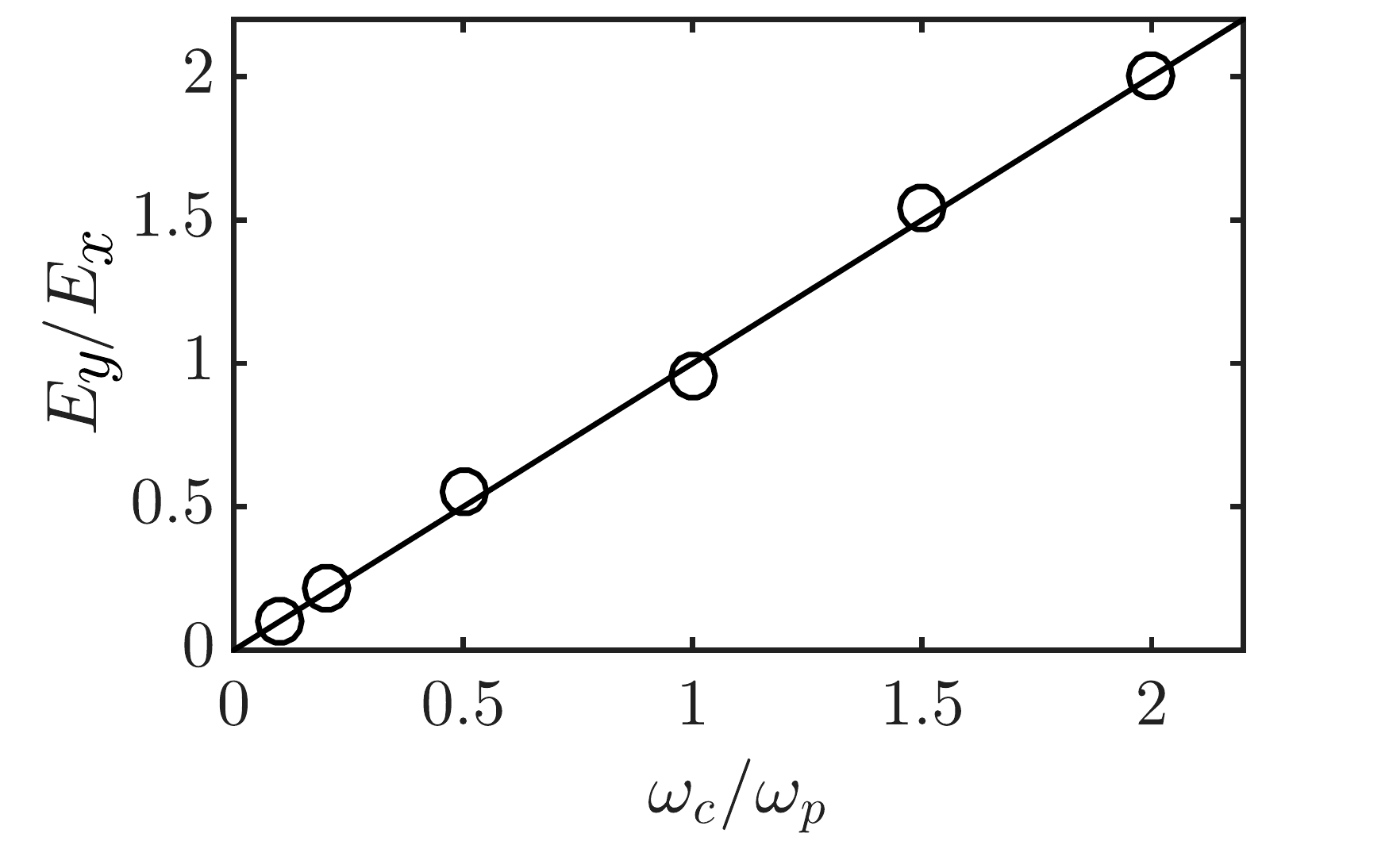}
\caption{Scaling of the ratio of the amplitude transverse component to the longitudinal component of the extraordinary plasma wave as a function of $\omega_c/\omega_p$. Simulations (circles) show good agreement with the theory (solid line) $|E_y/E_x| = \omega_c/\omega_p$}
\label{fig. xwaves}
\end{figure}

So far unmagnetized plasmas have been considered, which only allows exciting the branch of Langmuir waves. If the plasma is initially magnetized, and the direction of the photons is perpendicular to the magnetic field, the burst can excite the lower branch of the extraordinary modes \cite{Yoshii_PRL_1997} at $\omega = \omega_p$ or $\omega = kc$. It is straightforward to prove that in a 1D linear theory, the longitudinal electric field $E_x$ of the extraordinary mode, driven by the Compton Force, is identical to the one derived for the Langmuir branch, see Eq.(\ref{Eq:Efield_scaling}). Since $\omega = \omega_p$, one deduces \cite{Krall_Trivelpiece_1973} that the transverse (or electromagnetic) component of the electric field is $E_y = i(\omega_c/\omega_p)E_x$, where $i$ is the imaginary unit and $\omega_c = eB_0/mc$ the cyclotron frequency associated to the background magnetic field $B_0$. The ratio between the longitudinal and the transverse component of the electric field of the extraordinary mode has been accurately verified in our simulations, as seen in Fig.\ref{fig. xwaves}. Each run was carried out in a $24~d_e$ box with $\Delta x_1=0.01~d_e$ and $\Delta t =0.0099~\omega_p^{-1}$, with $256$ macro-particles per cell. The length of the radiation burst is $\lambda_p$. The energy of each photon and the energy density of the radiation are $\hbar\omega=50$ keV, and  $\mathcal{E}_0/\mathcal{E}_{\mathrm{min}}=30$ respectively.  

An important question to address is the conversion efficiency from the initial burst into extraordinary waves since the electromagnetic part of these modes can be transmitted to vacuum at the exit of the plasma. The incoming energy flux is on the order of $I_{\mathrm{in}} \sim \mathcal{E}_0c$ while the electromagnetic energy flux associated to the extraordinary waves is $I_{\mathrm{out}}\sim E_y^2c/4\pi$. Using Eq.(\ref{Eq:Efield_scaling}) for the amplitude of $E_x$, we obtain for a resonant driver (Eq.(\ref{Eq:Efield_scaling}) for $l=1$)
\begin{equation}
\eta=\frac{I_{\mathrm{out}}}{I_{\mathrm{in}}}\sim\frac{4\pi^3}{9}\left(\frac{\omega_c}{\omega_p}\right)^2\frac{r_e^3\mathcal{E}_0}{mc^2}.
\label{eq. XO eff}
\end{equation}

In the case of a low magnetization, $\omega_c/\omega_p \ll 1$, which implies that $\eta \ll 1$ since $r_e^3\mathcal{E}_0/mc^2 \sim (r_e/d_e)(\hbar\omega/mc^2) \ll 1$ for $\mathcal{E}_0 \gtrsim \mathcal{E_{\mathrm{min}}}$. The conversion efficiency $\eta$ cannot be above unity for an arbitrary values of $\mathcal{E}_0$. The electrostatic component of the extraordinary wave is limited to the wave-breaking limit \cite{Tajima_PRL_1979,Esarey_IEEE_1996} thus $\eta_{\mathrm{max}} \sim B_0^2/\mathcal{E}_0$. The power transmitted to vacuum will be further attenuated as the extraordinary mode must tunnel through the evanescent layer at the plasma/vacuum interface \cite{Yoshii_PRL_1997}. For high magnetization $\omega_c/\omega_p \gg 1$, there is no attenuation since the excited mode is allowed to propagate through the plasma/vacuum interface for any value of the decreasing plasma density (no evanescent layer). This stems from the hybrid frequency, which diminishes down the plasma gradient, still remaining greater than the plasma frequency where the mode was excited. Nonetheless, two factors can restrict the efficiency. First, when $E_y > mc\omega_p/e$, the motion of the electrons becomes relativistic and the classical scaling $|E_y/E_x| = \omega_c/\omega_p$ breaks. Secondly, if the background magnetic is sufficiently large, additional processes such as synchrotron emission \cite{LandauCTF} will impact the dynamics of the plasma electrons.

Current X-ray facilities are far from supplying the required energy density to observe these wakes in the laboratory. We present however a set of non ludicrous parameters for a possible future experiment: a plasma density $n_p= 10^{17}\mathrm{cm}^{-3}$, a non-resonant driver of length $L= \lambda_p/10$ (duration of 40 fs), and 100 eV X-rays, the required energy density would be $\mathcal{E} \simeq 3 \mathrm{mJ}/\mathrm{\mu m^3}$. The total energy of the driver would be $\mathcal{E} L^3 = 5$J. The amplitude of the wake would be on the order of MV/m, which can be measured experimentally.

In astrophysics, energetic photons emitted in extreme objects permeate tenuous plasmas. The spectrum of these photons encompasses a broad range of frequencies from optical to x-rays (or even gamma-rays). The total amount of energy released as well the energy density can be considerable, which counterbalance the low cross section associated to Compton scattering and therefore permit the generation of efficient plasma wakes. For example, extremely powerful sources/events e.g. $\gamma$-ray bursts (GRBs), can radiate, within seconds, more than $10^{50-54}$ erg, with a spectrum peak around $100$s of keV~\cite{Berger_ARAA_2014}. The photospheric radius, where most energy is radiated, amounts approximately to $R_{ph}\simeq10^{12-15}~\mathrm{cm}$~\cite{Daigne_MNRAS_2002}. Such bursts release a typical energy flux on the order of $10^{23}~\mathrm{erg/cm^{2}}s$ in the interstellar medium of density $n_p \sim 1 \mathrm{cm^{-3}}$. The energy density of the photon is on the order of $\mathcal{E}\sim 10^{13}~\mathrm{erg/ cm^3}$. These parameters satisfy  Eq.(\ref{eq:coherence_scale}) showing that Compton scattering may drive strong plasma wakes in the circumburst medium of GRBs.
X-ray bursts from magnetars show a radio counterpart which exhibit complex time-frequency structures similar to the ones recorded for the fast radio burst FRB 121102 \cite{Hessels_ApJL_2019,Mann_ApJL_2019}. In this scenario, the radio burst forms in a highly magnetized environment \cite{Chatterjee_Nat_2017}.
These generated X-waves may result in FRBs, which addresses one of the most important open questions  in astrophysics \cite{Platts_2018}. The efficiency of the present mechanism favorably scales linearly with $\sigma=\omega_c^2/\omega_p^2$, as others proposed processes, e.g. synchrotron maser emission from $e^+e^-$ collisionless shocks \cite{Beloborodov_arXiv_2019,Plotnikov_MNRAS_2019}, which scales as $1/\sigma$. 
To summarize, we have explored and unveiled a new regime of light-plasma interaction where photon bursts can excite efficiently plasma wakes via Compton Scattering. The plasma wakes correspond to the Langmuir and the extraordinary modes for unmagnetized and magnetized plasmas respectively, and can lead to the emission of radio waves in astrophysical scenarios.

\begin{acknowledgments}
 We would like to acknowledge useful discussions with Prof. Anatoly Spitkovsky. This work was supported by the European Research Council (ERC-2015-AdG Grant 695088), FCT (Portugal) grants PD/BD/114323/2016 in the framework of the Advanced Program in Plasma Science and Engineering (APPLAuSE, FCT grant No. PD/00505/2012). We acknowledge PRACE for awarding access to resource MareNostrum based in Spain. Simulations were performed at IST cluster (Portugal), and at MareNostrum (Spain).
\end{acknowledgments}

%\bibliography{paper_WFCS}% Produces the bibliography via BibTeX.

\begin{thebibliography}{100}
\bibitem{classical_field_LL}
The condition for a field  to be classical is, if the number of photons (of momentum $\hbar k$ ) in a volume $1/k^3$ is large compared with unity, i.e., $E^2c^3/\hbar\omega^4 \gg1$. This implies that low frequency fields are usually always classical, whereas very high frequency fields if sufficiently weak can never be classical \cite{LandauQED}.
\bibitem{field_amplitude}
For a laser ponderomotive drive, the amplitude of the wake is $E \sim en_p\lambda_p \propto (n_p)^{1/2}$. For a photon driver that only interacts via Compton scattering, only a fraction of the electrons are displaced. This fraction, being $(\nu_C / \omega_p)n_p$, adds a new factor that scales as $(n_p)^{-1/2}$, making the amplitude of the field independent of the plasma density.
\bibitem{threshold}
Wakes can probably be observed for $\nu_C/\omega_p < 1$. The streaming hot particles that are not caught back by the wave tend to produce electrostatic fluctuations of order $E_f$ \cite{Krall_Trivelpiece_1973}. This field results from the individual wakes of every streaming particles. A wake is observable if $E_{\mathrm{wake}} > E_f$, which corresponds to $\nu_C/\omega_p > g$, where $g = 1/n_p\lambda_{Ds}^3$ is the plasma parameter of the streaming hot particles with $\lambda_{Ds} \sim (\hbar\omega/ e^2n_s)^{1/2}$ and $n_s \sim n_p\nu_C/\omega_p$.
\end{thebibliography}

\end{document}